\documentclass[reqno,onecolumn,oneside]{paper}
\usepackage[english]{babel}
\usepackage{amsmath,amsthm,amsfonts,pifont,slashed,ifsym,yfonts,calligra,amssymb,latexsym,mathrsfs}
\usepackage[colorlinks]{hyperref}
\usepackage[all]{xy}
\usepackage[sort]{cite}

\newtheorem{theorem}{Theorem}[section]
\newtheorem{proposition}[theorem]{Proposition}
\newtheorem{lemma}[theorem]{Lemma}

\theoremstyle{definition}
\newtheorem{definition}[theorem]{Definition}
\newtheorem{exe}[theorem]{Example}

\theoremstyle{remark}

\newcommand{\op}{\operatorname{op}}
\newcommand{\edm}{\operatorname{End}_\QQ}
\newcommand{\obj}{\operatorname{Obj}}
\newcommand{\id}{\operatorname{id}}
\newcommand{\SL}{\mathcal{SL}}
\newcommand{\Q}{\mathcal{Q}}
\newcommand{\Edm}{\mathbf{End}_\QQ}
\newcommand{\MM}{\mathbf{M}}

\newcommand{\NN}{\mathbf{N}}
\newcommand{\LL}{\mathbf{L}}
\newcommand{\QQ}{\mathbf{Q}}
\newcommand{\RR}{\mathbf{R}}
\renewcommand{\hom}{\operatorname{Hom}}
\newcommand{\Hom}{\mathbf{Hom}}
\newcommand{\R}{\mathbb{R}}
\newcommand{\N}{\mathbb{N}}
\newcommand{\Z}{\mathbb{Z}}
\renewcommand{\wp}{\mathscr{P}}
\renewcommand{\phi}{\varphi}
\newcommand{\e}{\varepsilon}

\renewcommand{\d}{\delta}

\newcommand{\la}{\langle}
\newcommand{\ra}{\rangle}
\newcommand{\liff}{\Longleftrightarrow}

\newcommand{\under}{\backslash}

\newcommand{\lto}{\longrightarrow}
\newcommand{\To}{\Longrightarrow}
\newcommand{\lmapsto}{\longmapsto}
\newcommand{\ust}{\under_\star}
\newcommand{\ov}{\overline}

\begin{document}

\title{A Unified Algebraic Framework for Fuzzy Image Compression and Mathematical Morphology}

\author{Ciro Russo}

\institution{Dipartimento di Matematica e Informatica \\ University of Salerno, Italy \\ \texttt{cirusso@unisa.it}}

\maketitle

\begin{abstract}
In this paper we show how certain techniques of image processing, having different scopes, can be joined together under a common ``algebraic roof''.
\end{abstract}

\section*{Introduction}

In the last years, fuzzy logics and fuzzy set theory have been widely applied to image processing tasks. In particular, the theory of \emph{fuzzy relation equations}, deeply investigated in~\cite{dinolasessa}, is involved in many algorithms for compression and reconstruction of digital images (see, for example,~\cite{dinolarusso,hirotapedrycz,nobuharatakama}).

In such techniques, however, the approach is mainly experimental and the algebraic context is seldom clearly defined. Basically, most of the fuzzy algorithms for image compression, make use of join-product operators; after all, a complete lattice order and a multiplication that is residuated w.r.t. the lattice-order are the fundamental ingredients of these operators.

On the other hand, there exists another class of operators acting on digital images that, although having a completely different scope, has the same algebraic form: \emph{mathematical morphological operators}. Mathematical Morphology is a technique for image processing and analysis whose birth can be traced back to the book~\cite{matheron} by G. Matheron and whose establishment is due mainly to the work of Heijmans and Serra (e.g. \cite{goutsiasheijmans,heijmans,serra}). The basic problem in mathematical morphology is to design nonlinear operators that extract relevant topological or geometric information from images. This requires the development of a mathematical model for images and a rigorous theory that describes fundamental properties of the desirable image operators.

Essentially, mathematical morphological operators analyse the objects in an image by ``probing'' them with a small geometric ``model-shape'' (e.g., line segment, disc, square) called the \emph{structuring element}. These operators are defined on spaces having both a complete lattice order (set or fuzzy set inclusion, in concrete applications) and an external action from another ordered structure (the set of translations); more, they are usually coupled in adjoint pairs.

Regarding these different classes operators, what is really outstanding from an algebraic point of view is the fact that they can both be expressed in terms of a complete lattice order and a residuated product. Our aim is to show how all these operators can be joined together in a common mathematical context: the categories of \emph{quantale modules} and the operators called \emph{$\Q$-module transforms}. We will also show that such operators
\begin{itemize}
\item are precisely the $\Q$-module homomorphisms between free modules,
\item are completely determined by the mathematical counterpart of the coder (for compression algorithms) and of the structuring element (in the case of mathematical morphology).
\end{itemize}

Throughout the paper, due to space constraint, we will omit all the proofs of propositions and theorems; however they can all be found in \cite{russo}.

\section{Preliminaries}
\label{prel}

In this section we will briefly recall some basic notions on several ordered algebraic structures.

\begin{definition}\label{residuated map}
Let $\la X, \leq \ra$ and $\la Y, \leq \ra$ be two posets. A map $f: X \lto Y$ is said to be \emph{residuated} iff there exists a map $g: Y \lto X$ such that, for all $x \in X$ and for all $y \in Y$, $f(x) \leq y \ \iff \ x \leq g(y)$.

It is immediate to verify that the map $g$ is uniquely determined; we will call it the \emph{residual map} or the \emph{residuum} of $f$, and denote it by $f_*$. The pair $(f, f_*)$ is called \emph{adjoint}; a residuated map preserves all existing joins and its residuum preserves all existing meets.
\end{definition}

The category $\SL$ of sup-lattices is the one whose objects are complete lattices and morphisms are maps preserving arbitrary joins or, that is the same, residuated maps. For a sup-lattice $\LL$, we will use the notation $\LL = \la L, \vee, \bot \ra$. For any sup-lattice $\LL = \la L, \vee, \bot \ra$, it is possible to define a dual sup-lattice in an obvious way: if we consider the opposite partial order $\geq$, then $\LL^{\op} = \la L, \wedge, \top \ra$ is a sup-lattice and, clearly, $(\LL^{\op})^{\op} = \LL$.

\begin{proposition}\label{residprop}
Let $\la X, \leq \ra$ and $\la Y, \leq \ra$ be posets, and let $(f,f_*)$ be an adjoint pair, with $f: X \lto Y$. Then the following hold:
\begin{enumerate}
\item[$(i)$]$f$ is surjective \quad $\iff$ \quad $f_*$ is injective \quad $\iff$ \quad $f \circ f_* = \id_Y$;
\item[$(ii)$]$f$ is injective \quad $\iff$ \quad $f_*$ is surjective \quad $\iff$ \quad $f_* \circ f = \id_X$.
\end{enumerate}
\end{proposition}

A binary operation $\cdot$ on a partially ordered set $\la P, \leq \ra$ is said to be \emph{residuated} iff there exist binary operations $\under$ and $/$ on $P$ such that for all $x, y, z \in P$, $x \cdot y \leq z \ \textrm{iff} \ x \leq z/y \ \textrm{iff} \ y \leq x \under z$. The operations $\under$ and $/$ are referred to as the left and right \emph{residua} of $\cdot$, respectively. In other words, a residuated binary operation over $\la P, \leq \ra$ is a map from $P \times P$ to $P$ that is residuated in both arguments. In the situations where $\cdot$ is a monoid operation with a unit element $e$ and the partial order is a lattice order, we can add the monoid unit and the lattice operations to the similarity type to get an algebraic structure $\RR = \la R, \vee, \wedge, \cdot, \under, /, e \ra$ called \emph{residuated lattice}.

In the category $\Q$ of quantales, $\obj(\Q)$ is the class of complete residuated lattices and the morphisms are the maps preserving products, the unit, arbitrary joins and the bottom element. An alternative, yet equivalent, definition of quantale is the following

\begin{definition}\label{quantale}
A \emph{quantale}\index{Quantale} is an algebraic structure $\QQ = \la Q, \vee, \cdot, \bot, e \ra$ such that
\begin{enumerate}
\item[$(Q1)$]$\la Q, \vee, \bot \ra$ is a sup-lattice,
\item[$(Q2)$]$\la Q, \cdot, e \ra$ is a monoid,
\item[$(Q3)$]$x \cdot \bigvee\limits_{i \in I} y_i = \bigvee\limits_{i \in I} (x \cdot y_i)$ \ and \ $\left(\bigvee\limits_{i \in I} y_i\right) \cdot x = \bigvee\limits_{i \in I} (y_i \cdot x)$ \ for all $x \in Q$, $\{y_i\}_{i \in I} \subseteq Q$.
\end{enumerate}
$\QQ$ is said to be \emph{commutative} if so is the multiplication. Obviously, if $\QQ$ is commutative then the two residua coincide and $x/y = y \under x$ is denoted by $y \to x$.
\end{definition}

Before giving examples of quantale structures interesting for the scope of this paper, we recall that a binary operation $\ast: [0,1]^2 \lto [0,1]$ is called a \emph{triangular norm}, \emph{t-norm} for short, provided it is associative, commutative, monotone in both arguments and has $1$ as the neutral element. A t-norm $\ast$ is called \emph{left-continuous} if, for all $\{x_n\}_{n \in \N}, \{y_n\}_{n \in \N} \in [0,1]^\N$,
\begin{center}
$\left(\bigvee_{n \in \N} x_n\right) \ast \left(\bigvee_{n \in \N} y_n\right) = \bigvee_{n \in \N} (x_n \ast y_n).$
\end{center}
In this case, clearly, $\ast$ is residuated and its residuum (unique, since $\ast$ is commutative) is given by \ $x \to y = \bigvee\{z \in [0,1] \mid z \ast x \leq y\}$.
\begin{exe}
If $\ast$ is any left-continuous t-norm on the real unit interval, then $\la [0,1], \vee, \ast, 0, 1 \ra$ is a commutative quantale.
\end{exe}

\section{Join-product operators in Image Processing}
\label{fuz}

\subsection{Fuzzy algorithms for image compression and reconstruction}
\label{fuzalgsec}

In the literature of image compression, the fuzzy approach is based essentially on the theory of fuzzy relation equations. The underlying idea is the following: a grey-scale image is a matrix in which every element represents a pixel and its value, included in the set $\{0, \ldots, 255\}$ in the case of a 256-bit encoding, is the grey-level. Then, if we normalize the set $\{0, \ldots, 255\}$ by dividing each element by $255$, grey-scale images can be modeled equivalently as fuzzy relations, fuzzy functions (i.e. $[0,1]$-valued maps), fuzzy subsets of a given set or $[0,1]$-valued matrices. A similar model is used for RGB colour images, where each image is represented by three fuzzy relations (respectively: functions, sets or matrices).

So, if we consider a grey-scale image $I$ of sizes $m \times n$ ($m, n \in \N$), we can see it as an $m \times n$ matrix $I_{ij}$ whose values are in $[0,1]$. Now we consider two natural numbers $a \leq m$ and $b \leq n$ and fix a $[0,1]$-valued map in four variables $C \in [0,1]^{m \times n \times a \times b}$ --- usually called \emph{coder} or \emph{codebook}; then we compress the image $I$ into an image $I' = I'_{hk}$ of sizes $a \times b$ by setting
\begin{equation}\label{com}
I'_{hk} = \bigvee_{i,j} I_{ij} \ast C_{ijhk},
\end{equation}
where $\ast$ is any left-continuous t-norm on $[0,1]$. The reconstructed image $I'' = I''_{ij}$ is defined by
\begin{equation}\label{rec}
I''_{ij} = \bigwedge_{h,k} C_{ijhk} \to_\ast I'_{hk},
\end{equation}
where $\to_\ast$ is the residuum of $\ast$.

Even if some fuzzy algorithms for image compression may look different at a first glance, most of them can be rewritten in a form similar to (\ref{com}), with a reconstruction process that will consequently look like (\ref{rec}).

\subsection{Dilation and erosion in Mathematical Morphology}
\label{mathmorphsec}

In Mathematical Morphology binary images are modeled, in the wake of tradition and intuition, as subspaces or subsets of a suitable space $E$, which is assumed to possess some additional structure (topological space, metric space, graph, etc.), usually depending on the kind of task at hand.

Concretely, the class of $n$-dimensional binary images is represented as $\wp(E)$, where $E$ is, in general, $\R^n$ or $\Z^n$. In the first case we have continuous binary images, otherwise we are dealing with discrete binary images. The basic relations and operations between images of this type are essentially those between sets, namely set inclusion, union, intersection and complementation. It is intuitively clear, then, that complete lattices are the algebraic structures required for abstracting the ideas introduced so far. 

\begin{definition}\label{dilero}
Let $\LL$, $\MM$ be complete lattices.  A map $\d: L \lto M$ is called a \emph{dilation} if it distributes over arbitrary joins: $\d\left({}^\LL\bigvee X \right) = {}^\MM\bigvee \d(X)$, for every $X \subseteq L$. A map $\e: M \lto L$ is called an \emph{erosion} if it distributes over arbitrary meets: $\e\left({}^\MM\bigwedge Y \right) = {}^\LL\bigwedge \e(Y)$, for every $Y \subseteq M$.

Two maps $\d: L \lto M$ and $\e: M \lto L$ are said to form an \emph{adjunction}, $(\d,\e)$, between $\LL$ and $\MM$ if $\d(x) \leq y \liff x \leq \e(y)$, for all $x \in L$ and $y \in M$.\footnote{Notice that the notation used in the literature of Mathematical Morphology is slightly different. Indeed, an adjunction is presented with a dilation in the second coordinate and an erosion in the first. Here we use such a reversed notation because, as we will see, adjunctions are adjoint pairs in the sense of Definition \ref{residuated map}.}
\end{definition}
Assume that $\d: \LL \lto \MM$ is a dilation. For $x \in L$, we can write \ $\d(x) = \bigvee_{y \leq x} \d(y)$, \ where we have used the fact that $\d$ distributes over join. Every dilation defined on $\LL$ is of the form above, and the adjoint erosion is given by \ $\e(y) = \bigvee_{\d(x) \leq y} x$.

Now, keeping in mind the models $\R^n$ and $\Z^n$, it is possible to introduce the concepts of \emph{translation} of an image and \emph{translation invariance} of an operator, by means of the algebraic operation of sum. Indeed, let $E$ be $\R^n$ or $\Z^n$ and consider the complete lattice $\wp(E)$; given an element $h \in E$, we define the $h$-\emph{translation} $\tau_h$ on $\wp(E)$ by setting, for all $X \in \wp(E)$, $\tau_h(X) = X + h = \{x + h \mid x \in X\}$, where the sum is intended to be defined coordinatewise.

Next, we consider the case of operators that are translation invariant; here the sets $\d(\{x\})$ are translates of a fixed set, called the \emph{structuring element}, by $\{x\}$. An operator $f: \wp(E) \lto \wp(E)$ is called \emph{translation invariant}, \emph{T-invariant} for short, if $\tau_h \circ f = f \circ \tau_h$ for all $h \in E$. It is proved in \cite{heijmans} that every T-invariant dilation on $\wp(E)$ is given by $\d_A(X) = \bigcup_{x \in X} A + x$, and every T-invariant erosion is given by $\e_A(X) = \{y \in E \mid A + y \subseteq X\} = \{y \in E \mid y \in X + \breve A\}$, where $A$ is an element of $\wp(E)$, called the \emph{structuring element}, and $\breve A = \{- a \mid a \in A\}$ is the reflection of $A$ around the origin.

Now we observe that the above expressions for erosion and dilation can also be written, respectively, as
\begin{equation}\label{trinv}
\d_A(X)(y) = \bigvee_{x \in E} A(y - x) \wedge X(x), \qquad \e_A(Y)(x) = \bigwedge_{y \in E} A(y - x) \to Y(y),
\end{equation}
where each subset $X$ of $E$ is identified with its (Boolean) membership function and $X \to Y =: X^c \vee Y$.
Moving from these expressions, and recalling that $\wedge$ is a residuated commutative operation (that is, a continuous t-norm) whose residuum is $\to$, it is possible to extend these operations from the complete lattice of sets $\wp(E) = \{0,1\}^E$ to the complete lattice of fuzzy sets $[0,1]^E$, by means of left-continuous t-norms and their residua. What we do, concretely, is to extend the morphological image operators of dilation and erosion, from the case of binary images, to the case of grey-scale images.

So let $\ast$ be a left-continuous t-norm and $\to$ be its residuum; a grey-scale image $X$ is a fuzzy subset of $E$. Given a fuzzy subset $A \in [0,1]^E$, called a \emph{fuzzy structuring element}, the operators
\begin{center}$
\d_A(X)(y) = \bigvee_{x \in E} A(y - x) \ast X(x), \qquad \e_A(X)(x) = \bigwedge_{y \in E} A(y - x) \to X(y)
$\end{center}
are, respectively, a translation invariant dilation and erosion on $[0,1]^E$.

\section{Quantale modules}
\label{qmod}

\begin{definition}\label{modules}
Let $\QQ$ be a quantale and $\MM = \la M, \vee, \bot \ra$ a sup-lattice. $\MM$ is a (left) \emph{$\QQ$-module} if there exists an external binary operation, called \emph{scalar multiplication}, \ $\star: (q,m) \in Q \times M \lmapsto q \star m \in M$, \ such that
\begin{enumerate}
\item[$(M1)$]$(q_1 \cdot q_2) \star m = q_1 \star (q_2 \star m)$, for all $q_1, q_2 \in Q$ and $m \in M$;
\item[$(M2)$]the external product is distributive with respect to arbitrary joins in both coordinates or --- that is the same --- it is residuated;
\item[$(M3)$]$e \star m = m$.
\end{enumerate}
\end{definition}
From $(M2)$ it follows that, for all $q \in Q$, there exists the residual map $(q^\star)_*$ of $q^\star$, and for all $m \in M$ there exists the residual map $(^\star m)_*$ of $^\star m$. In particular it is possible to define another external operation over $M$:
\begin{center}
$\ust: (q,m) \in Q \times M \lmapsto q \ust m = (q^\star)_*(m) \in M.$
\end{center}

\begin{exe}\label{freemodule}
Let $\QQ$ be a quantale and $X$ be an arbitrary non-empty set. We can consider the sup-lattice $\QQ^X = \la Q^X, \vee^X, \bot^X \ra$, where $\bot^X$ is the $\bot$-constant function from $X$ to $Q$ and the join and the scalar multiplication $\star$ are defined pointwisely from those in $\QQ$.

Then $\QQ^X$ is a left $\QQ$-module and, for all $q \in Q$, $f \in Q^X$ and $x \in X$, $(q \ust f)(x) = q \under f(x)$.
\end{exe}
It is easy to show that the module in the previous example is the free $\QQ$-module over the set of generators $X$.

Definition, and properties, of right $\QQ$-modules are completely analogous. If $\QQ$ is commutative, right and left $\QQ$-modules coincide and we will say simply $\QQ$-modules. If $\QQ$ is a quantale and $\MM$ is a left $\QQ$-module, the dual sup-lattice $\MM^{\op}$ is a right $\QQ$-module (and vice versa) with the external multiplication $\ust$.\footnote{In what follows, in all the definitions and results that can be stated both for left and right modules, we will refer generically to ``modules''~--- without specifying left or right~--- and we will use the notations of left modules.}

Let $\QQ$ be a quantale and $\MM_1, \MM_2$ be two $\QQ$-modules. A map $f: M_1 \lto M_2$ is a $\QQ$-module homomorphism if $f\left({}^{\MM_1}\bigvee_{i \in I} m_i\right) = {}^{\MM_2}\bigvee_{i \in I} f(m_i)$ for any family $\{m_i\}_{i \in I} \subseteq \MM_1$, and $f(q \star_1 m) = q \star_2 f(m)$, for all $q \in Q$ and $m \in M_1$, where $\star_i$ is the external product of $\MM_i$, for $i = 1,2$.

\begin{proposition}\label{dualhomomq}
Let $\QQ$ be a quantale, $\MM_1$, $\MM_2$ be two $\QQ$-modules and $f: \MM_1 \lto \MM_2$ be a homomorphism. Then $f$ is a residuated map and the residual map $f_*: M_2 \lto M_1$ is a $\QQ$-module homomorphism between $\MM_2^{\op}$ and $\MM_1^{\op}$.
\end{proposition}

\begin{definition}\label{hommq}
Let $\MM$ and $\NN$ be two $\Q$-modules and $\hom_\QQ(\MM,\NN)$, the set of all the homomorphisms from $\MM$ to $\NN$. Then the structure $\Hom_\QQ(\MM,\NN) = \la \hom_\QQ(\MM,\NN), \sqcup, \bot^\bot \ra$, with the pointwise join and the $\bot$-constant homomorphism as bottom element, is a sup-lattice; moreover, if $\QQ$ is a commutative quantale, $\Hom_\QQ(\MM,\NN)$ is a $\QQ$-module with the scalar multiplication $\diamond$ defined, again, pointwisely: for all $q \in Q$ and $h \in \hom_\QQ(\MM,\NN)$, $q \diamond h$ is the homomorphism defined by $(q \diamond h)(x) = q \star h(x) = h(q \star x)$, for all $x \in M$.

If $\NN = \MM$, $\Hom_\QQ(\MM,\MM)$ is denoted by $\Edm(\MM) = \la \edm(\MM), \sqcup, \bot^\bot \ra$.
\end{definition}

\section{Quantale module transforms}
\label{mqtransec}

In this section we introduce the $\Q$-module transforms and we list some results about them. Then we present a classification of these operators that have interesting theoretical and concrete consequences. For an extensive treatment of $\Q$-module transforms, the reader may refer to \cite{russo}.

\begin{definition}\label{qwtransform}
Let $\QQ \in \Q$ and $X, Y$ be non-empty sets and let us consider the free $\QQ$-modules $\QQ^X$ and $\QQ^Y$. We will call a \emph{$\Q$-module transform} between $\QQ^X$ and $\QQ^Y$, with \emph{kernel} $p$, the operator $H_p: Q^X \lto Q^Y$ defined by
\begin{center}
$H_p f(y) = \bigvee_{x \in X} f(x) \cdot p(x,y) \quad \textrm{for all } y \in Y,$
\end{center}
where $p \in Q^{X \times Y}$. Its \emph{inverse transform} $\Lambda_p: Q^Y \lto Q^X$ is the map defined by
\begin{center}
$\Lambda_p g(x) = \bigwedge_{y \in Y} g(y) / p(x,y) \quad \textrm{for all } x \in X.$
\end{center}
\end{definition}

\begin{theorem}\label{wadjointpair}
Let $\QQ \in \Q$, $X, Y$ be two non-empty sets and $p \in Q^{X \times Y}$. If $H_p$ is the $\Q$-module transform, with kernel $p$, between $\QQ^X$ and $\QQ^Y$, and $\Lambda_p$ is its inverse transform, then the following hold:
\begin{enumerate}
\item[$(i)$]$(H_p,\Lambda_p)$ is an adjoint pair, i.e. $H_p$ is a residuated map and $\Lambda_p = {H_p}_*$;
\item[$(ii)$]$H_p \in \hom_\QQ\left(\QQ^X,\QQ^Y\right)$ and $\Lambda_p \in \hom_\QQ\left(\left(\QQ^Y\right)^{\op},\left(\QQ^X\right)^{\op}\right)$.
\end{enumerate} 
\end{theorem}

The following classification of the kernels has a few interesting theoretical implications but it is important for applications to image processing. We refer to \cite{dinolarusso} (where an orthonormal transform is presented), \cite{russothesis} and \cite{russo} for details.
\begin{definition}\label{coders}
Let $\QQ \in \Q$, and $X, Y$ be non-empty sets. Let us consider a map $p \in Q^{X \times Y}$; we set the following definitions:
\begin{enumerate}
\item[$(i)$]$p$ is called a \emph{coder} iff there exists an injective map $\e: Y \lto X$ such that $e \leq p(\e(y),y)$ for all $y \in Y$;
\item[$(ii)$]$p$ is said to be \emph{normal} iff there exists an injective map $\e: Y \lto X$ such that $p(\e(y),y) = e$ for all $y \in Y$;
\item[$(iii)$]$p$ is said to be \emph{strong} iff it is normal and \ $p(\e(y_1),y_2) = \bot$ \ for all $y_1 \neq y_2 \in Y$;
\item[$(iv)$]$p$ is said to be \emph{orthogonal} iff $p(x,y_1) \cdot p(x,y_2) = \bot$ for all $y_1, y_2 \in Y$ such that $y_1 \neq y_2$ and for all $x \in X$;
\item[$(v)$]$p$ is said to be \emph{orthonormal} iff it is orthogonal and normal.
\end{enumerate}
If $p$ is a coder, the $\Q$-module transform $H_p$ is called \emph{faithful} and, if $p$ is normal, strong, orthogonal or orthonormal, the corresponding transform will have the same adjective. Also, we observe that $(v) \To (iii) \To (ii) \To (i)$.
\end{definition}

\begin{theorem}\label{sadjointpair}
Let $\QQ \in \Q$ and let $H_p$ be a $\Q$-module strong transform, by the coder $p \in Q^{X \times Y}$, with inverse transform $\Lambda_p$. Then \ $H_p \circ \Lambda_p = \id_{Q^Y}$; \ thus $H_p$ is onto and, by Proposition~\ref{residprop}, $\Lambda_p$ is one-one.
\end{theorem}

\begin{lemma}\label{transf=coder}
Let $\QQ \in \Q$, $X$ be a non-empty set, $Y$ be a non-empty subset of $X$ and $p, p' \in Q^{X \times Y}$ be two maps. Then $H_p = H_{p'}$ if and only if $p = p'$.
\end{lemma}

The previous result ensures us that a $\Q$-module transform $H_p$ is completely determined by its kernel $p$, while next result is the converse of Theorem~\ref{wadjointpair}$(ii)$; it proves that all the homomorphisms between free modules are transforms.

\begin{theorem}\label{repr}
The sup-lattices $\Hom_\QQ(\QQ^X,\QQ^Y)$ and $\QQ^{X \times Y}$ are isomorphic. And, if $\QQ$ is commutative, they are isomorphic also as $\QQ$-modules. In particular $\Edm(\QQ^X) \cong \QQ^{X \times X}$.
\end{theorem}

Here we just ``scratched the surface'' of $\Q$-module transforms, especially in order to focus the attention on the main thesis of the present paper, i.e. the fact that fuzzy image compression and mathematical morphological operators fall within the same class of operators under an algebraic point of view. We, again, refer to \cite{russothesis} or \cite{russo} the reader who may be interested in $\Q$-module transforms.

\section{Conclusion}
\label{conc}

As the reader may have already noticed, the operators in Section \ref{fuz} are all special cases of $[0,1]$-module transforms. We will now analyse them in detail.

Let us consider the compression operator defined in Subsection \ref{fuzalgsec}. Its domain is $[0,1]^{m \times n}$ and its codomain is $[0,1]^{a \times b}$; in the light of the definitions and results presented so far, we get immediately that (\ref{com}) is the $[0,1]$-module transform $H_C: [0,1]^{m \times n} \lto [0,1]^{a \times b}$ with kernel $C \in [0,1]^{m \times n \times a \times b}$ and the reconstruction (\ref{rec}) is its inverse transform $\Lambda_C: [0,1]^{a \times b} \lto [0,1]^{m \times n}.$

We already observed in Subsection~\ref{mathmorphsec} that dilations are precisely the sup-lattice homomorphisms while erosions are their residua. In order to faithfully represent dilations and erosions that are translation invariant as $\Q$-module transforms from a free $[0,1]$-module to itself, we make the further assumption that the set over which the free module is defined has the additional structure of Abelian group. So let $\mathbf{X} = \la X, +, -, 0 \ra$ be an Abelian group, $\ast$ a t-norm on $[0,1]$ and consider the free $[0,1]$-module $[0,1]^X$. For any element $k \in [0,1]^X$, we define the two variable map $\ov k: (x,y) \in X \times X \lmapsto k(y - x) \in [0,1]$. Then, for all $k \in [0,1]^X$, the translation invariant dilation, on $[0,1]^X$, whose structuring element is $k$, is precisely the $\Q$-module transform $H_{\ov k}: [0,1]^X \lto [0,1]^X$, with the kernel $\ov k$ defined above. Obviously, the translation invariant erosion whose structuring element is $k$ is $\Lambda_{\ov k}$.

Then the representation of both classes of operators as $\Q$-module transforms is rather trivial. Actually, what we want to point out here is that, if we drop the assumption that our quantale is defined on $[0,1]$, the classes of transforms defined in this section become much wider. The purpose of this consideration is not to suggest purely speculative abstractions but, rather, to underline that suitable generalizations of these operators exist already and they may be useful provided their underlying ideas are extended to tasks involving other quantales.


\begin{thebibliography}{99}

\bibitem{dinolarusso}
Di Nola A., Russo C., {\em \L ukasiewicz Transform and its application to compression and reconstruction of digital images}, Informat. Sci., {\bf 177}(6), 1481--1498, 2007

\bibitem{dinolasessa}
Di Nola A., Sessa S., Pedrycz W., Sanchez E., {\em Fuzzy relation equations and their applications to knowledge engineering}, Kluwer, Dordrecht, 1989

\bibitem{goutsiasheijmans}
Goutsias J., Heijmans H.J.A.M., {\em Fundamenta Morphologicae Mathematicae}, Fundamenta Informaticae, {\bf 41}, 1--31, 2000

\bibitem{heijmans}
Heijmans H.J.A.M., {\em Morphological Image Operators}, Ac. Press, Bos\-ton, 1994

\bibitem{hirotapedrycz}
Hirota K., Pedrycz W., {\em Fuzzy relational compression}, IEEE Trans. Syst. Man Cyber. -- Part B, {\bf 29}(3), 407--415, 1999

\bibitem{matheron}
Matheron G., {\em Random Sets and Integral Geometry}, J. Wiley \& Sons, N.Y., 1975

\bibitem{nobuharatakama}
Nobuhara H., Takama Y., Hirota K., {\em Image compression/reconstruction based on various types of fuzzy relational equations}, Trans. Inst. Elec.l Eng. Japan, {\bf 121-C}(6), 1102--1113, 2001

\bibitem{russothesis}
Russo C., {\em Quantale Modules, with Applications to Logic and Image Processing}, Ph.D. Thesis, University of Salerno, Italy, 2007

\bibitem{russo}
Russo C., {\em Quantale Modules and their Operators, with Applications}, J. Logic Comput., doi:10.1093/logcom/exn088, 2008

\bibitem{serra}
Serra J., {\em Image Analysis and Mathematical Morphology}, Ac. Press, London, 1982

\end{thebibliography}
\end{document}